# Thin-film PMUTs: A Review of 40 Years of Research

Kaustav Roy[a,b], Joshua En-Yuan Lee[c] and Chengkuo Lee[a,b]*

[a]Department of Electrical and Computer Engineering, National University of Singapore, Singapore 117583
[b]Center for Intelligent Sensors and MEMS (CISM), National University of Singapore, Singapore 117608
[c]Institute of Microelectronics, A*STAR, Singapore 117685

*All correspondence should be addressed to elelc@nus.edu.sg*

**Abstract**
A concise review on the thin-film PMUTs, which has been one of the rather important research topics amongst micro-ultrasound experts is being reported. It has been rigorously surveyed, scrutinized, and perceived that the work in this direction began nearly 44 years ago, with the primitive development of functional piezoelectric thin-film material, and now there are already three major companies commercializing them on a bulk scale. This fascinating fact illustrates the enormous contribution made by more than 70 different centers, research institutes, and agencies spread across 4 different continents to develop the vast know-how of these devices' design, make, and function. This review covers such important contributions in a short but comprehensive fashion and in particular, intends to educate the readers about – the global scenario of PMUTs, the principles governing their design, the ways following their make, all non-conventional useful PMUT designs, and lastly, category wise applications. Crucial comparison charts in terms of thin-film piezoelectric material used in PMUTs, and in terms of targeted applications are also depicted and discussed, which is believed to enlighten any MEMS designer planning to begin working with PMUTs. Moreover, each relevant section is provided with a crisp future forecast, from the author's past knowledge and expertise in this very field of research along with the elements of a careful literature survey. In short, this review can be considered a one-stop time-efficient guide for whoever is interested in knowing about these small devices.

## Introduction

The field of transducers has undergone several revolutions. One of those reforms was the creation of nanotechnology, which enabled the making of very small transducers, in the range of sub-micron dimensions. This along with the development of material research gave birth to microelectromechanical systems (MEMS). Soon the macro world of sensors shifted to micro leading to an ever-increasing demand in the field. Today, most smartphones have MEMS inertial systems inside. MEMS has also revolutionized the field of ultrasound, leading to the creation of tiny ultrasound transducers. These are popularly known as Micromachined Ultrasound Transducers (MUTs) and can be broadly classified into – Capacitive MUTs [1] (CMUTs) and Piezoelectric MUTs (PMUTs).

The key component of a MUT is a suspended micro-diaphragm. A CMUT's diaphragm consists of either a single layer of a conducting material or a nonconducting material with an electrode layer on it, suspended generally with a 0.5 to 2 μm gap from the grounded substrate which is electrostatically actuated with an alternating electric field (AC). Alternatively, a PMUT's diaphragm consists of at least four layers — a passive layer, and a piezoelectric layer sandwiched between metal electrodes, which can be dynamically actuated with AC.

Although there have been existing reviews on PMUTs [2], [3], which provide an understanding of the subject, still, most of the important information remains concealed. Some of such are – (a) the historical evolution of PMUTs, notable PMUT hotspots, publication statistics, and PMUT commercialization, (b) PMUT's working principle and design maps, (c) special PMUTs with enhanced capabilities – structurally modified PMUTs and flexible/stretchable PMUTs and (d) applications of PMUTs specifically as transmitters, receivers, and transceivers. Above all, an important definition of PMUTs in terms of the thickness of the piezoelectric layer is missing in either of the reviews without which it is difficult to classify PMUTs into thin or thick film categories. This review is thus prepared to address all such missing information and serves to provide a holistic overview of the past, the current research in the very direction, and future predictions. It starts with the PMUT evolution, followed by PMUT design basics, novel piezoelectric thin-films along with methods to fabricate PMUTs. It subsequently continues explaining special PMUTs with unconventional designs and their research implications followed by a vivid description of the major applications using PMUTs for constructing a transmitter, a receiver, or a transceiver. Two critical comparison charts about various thin-film piezoelectric materials and PMUT applications have also been tabulated, which can easily give the reader a feel for the numbers.

## The PMUT Evolution: Past to Present
### About Thin-film PMUTs
#### What are PMUTs?

PMUTs are micro-membrane ultrasound transducers that are backed by an acoustic cavity, fabricated using the VLSI nanofabrication techniques, and can be made in various shapes and dimensions. The transduction mechanism in PMUTs is governed by thin-film piezoelectricity that makes them vibrate. PMUTs are generally operated at their resonant frequency for obtaining response and can function as a transmitter/actuator, receiver/sensor, or as transceiver. As a transducer, PMUTs have three advantages over the commercially available bulk ultrasound transducers: (a) They require lower actuation voltages to produce a unit acoustic pressure (Pa/V/mm$^2$), making them suitable for portable low-power applications (b) They can operate both in air and water due to an efficient impedance matching with the surrounding medium, and (c) They can be made in tiny form factors, thereby enhancing suitability for space-constrained applications. Figure 1(a(i)) represents a PMUT die with elements arranged in 2D m x n arrays. One of the elements in the die is cut (BB') in the form of a three-quarter cross-section and is depicted in Figure 1(a(ii)). It shows all the layer constituents of a typical PMUT fabricated by bulk micromachining of silicon. Figure 1(a(iii)) shows a portion (AA') of a linear PMUT array. Figure 1(b) shows a collage of different PMUT chips fabricated by the author's groups all having their unique application functionalities.

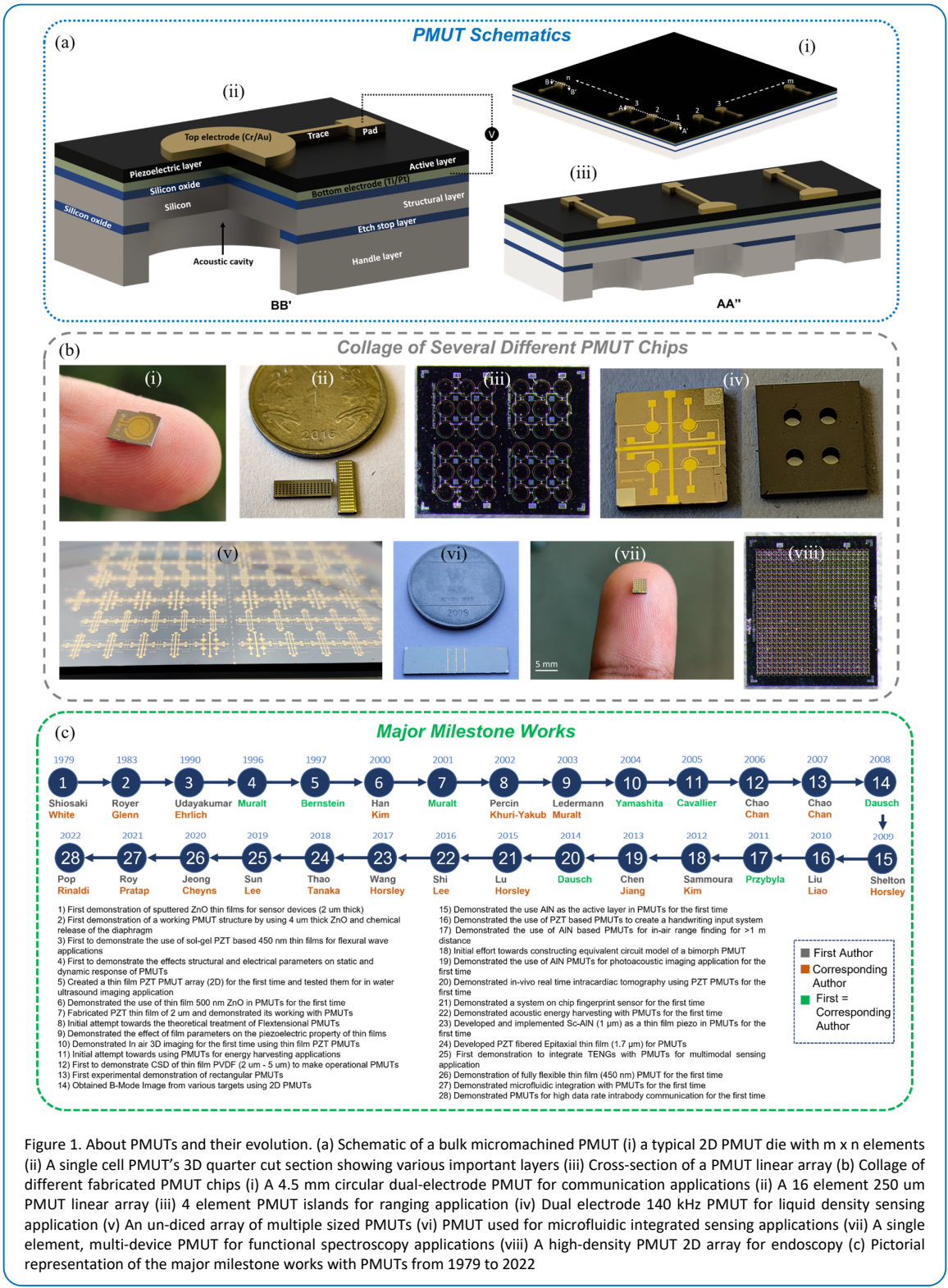

Figure 1. About PMUTs and their evolution. (a) Schematic of a bulk micromachined PMUT (i) a typical 2D PMUT die with m x n elements (ii) A single cell PMUT's 3D quarter cut section showing various important layers (iii) Cross-section of a PMUT linear array (b) Collage of different fabricated PMUT chips (i) A 4.5 mm circular dual-electrode PMUT for communication applications (ii) A 16 element 250 um PMUT linear array (iii) 4 element PMUT islands for ranging application (iv) Dual electrode 140 kHz PMUT for liquid density sensing application (v) An un-diced array of multiple sized PMUTs (vi) PMUT used for microfluidic integrated sensing applications (vii) A single element, multi-device PMUT for functional spectroscopy applications (viii) A high-density PMUT 2D array for endoscopy (c) Pictorial representation of the major milestone works with PMUTs from 1979 to 2022

**Piezoelectric Active Layer Thickness in Thin-film PMUTs**

PMUTs function on the lateral dynamic strain developed in the piezoelectric thin-film and perform best if the thickness of the film is below a critical limit, above which the transverse strain cross-interference becomes noticeable. There are works that define this limit where the thickness has been experimentally found to be 3 µm in a PZT {100}-textured thin-film [4], above which there has been a significant decay in the lateral thin-film based ($31$,f) piezoelectric coefficient. Thus, in this review, only relevant articles that satisfy this criterion have been included.

**Historical Development**

The first spark towards PMUT creation dates to the year 1979 when Shiosaki et al. [5] developed thin-film sputtered Zinc Oxide (ZnO) for sensors. Four years later, Royer et al. [6] demonstrated a working PMUT structure by using thin-film ZnO with the chemical release of PMUT diaphragm. In 1990, Udayakumar et al. [7] used sol-gel Lead Zirconate Titanate (PZT) for thin-film for flexural wave applications followed by Muralt et al. demonstrating the effects of structural and electrical parameters on PMUT's response in 1996 [8]. Next, Bernstein [9] in 1997, created a thin-film PZT PMUT array for in-water ultrasound imaging. In 2000, Han et al. [10] fabricated ZnO film of 500 nm thickness in PMUTs for the first time which was the thinnest of all films till then. In 2001, Muralt [11] fabricated a 2 µm thin-film with PMUTs, followed by Percin et al. [12] attempting towards theoretical treatment of flextensional PMUTs in 2002. In 2003, Ledermann et al. [4] demonstrated the effect of film parameters on the piezoelectricity of thin-films followed by Yamashita demonstrated in-air 3D imaging for the first time using thin-film PZT PMUTs [13] in 2004. Next, Dogheche et al. [14] demonstrated energy harvesting applications using PMUTs in 2005. In 2006, Chao et al. [15] demonstrated chemical solution deposition (CSD) of thin-film PVDF to make operational PMUTs. This was the first-time polymer piezoelectric material was used in PMUTs. Next, in 2007, Chao et al. reported on rectangular PMUTs [16] followed by Dausch obtaining B-mode ultrasound images from various targets using a 2D PMUT array [17] in 2008. Next, in 2009, Shelton first proposed the use of Aluminum Nitride (AlN) thin-film as the active layer in PMUTs [18], leading to the creation of CMOS-compatible PMUT arrays. This work was followed by Liu et al. proposing a handwriting input system using PZT-based PMUTs [19], in 2010. Subsequently, Przybyla [20] reported on the use of AlN PMUTs for in-air range finding for > 1 m distance in 2011. Next, in 2012 there was an initial effort toward constructing an equivalent circuit model of a bimorph PMUT from Sammoura et al. [21]. In 2013, Chen et al. demonstrated the use of AlN PMUTs for photoacoustic imaging applications for the first time [22]. Next, Dausch demonstrated in-vivo real-time intracardiac tomography using PZT PMUTs in 2014 [23] which is considered as the major in-body medical ultrasound imaging application demonstrated till now. In 2015, Lu et al. [24] demonstrated a system-on-chip fingerprint sensor, which is one of the most important applications having relevance to consumer electronics. Next, in the year 2016, Shi et al. [25] demonstrated acoustic energy harvesting with PMUTs comprehensively, followed by Wang et al. developing the Scandium-doped AlN (Sc-AlN) for PMUTs for the first time in 2017 [26]. Subsequently, in 2018, Thao et al. developed PZT-fibered epitaxial thin-film PMUTs [27], which was followed by the development of energy-efficient Triboelectric Nanogenerator (TENG) powered PMUT for multimodal sensing applications by Sun et al. [28] in 2019. In 2020, Jeong et al. [29] developed a fabrication flow for flexible thin-film PMUTs, which has relevance in medical wearables. In the year 2021, Roy et al. [30] developed a microfluidic integrated platform with PMUTs and demonstrated fluid parameter sensing, followed by applying PMUTs for high data rate intrabody communication by Pop et al. for the first time in 2022 [31]. Thus, looking at history, there has been a huge progress in terms of material development, innovative PMUT designs and applications. The pictorial representation of the milestone works is depicted in Figure 1(c).

**PMUT Hotspots and Publication Statistics**

Literature reveals nearly 70 places worldwide, across 4 continents (in terms of academic universities and agencies/companies) working on PMUTs. Asia has 30, Europe has 18, North America has 21 and Australia has 1 PMUT hotspot respectively (Figure 2(a)). Publication-wise, the number of research articles has increased exponentially since 2012 with 14 published articles (Figure 2(b)) to 79 in 2021 which indicates an ever-increasing demand for researchers to work in this domain to create several novelties.

**PMUT Commercialization**

PMUTs have been commercialized by three companies (Figure 2(c)) – Qualcomm, TDK, and Exo. Qualcomm has developed the first commercial in-display PMUTs (called 3D sonic sensor) to map 3D fingerprints. TDK has commercialized Application Specific Integrated Circuits (ASICs) bonded PMUTs (CH101, CH201) suitable for range finding till 1.2 m and 5 m respectively. Exo has developed a handheld prototype – Cello containing 4096 low-powered PMUTs for multi-harmonic imaging.

**Author's Sectional Forecast**

There has been rapid growth in the number of hotspots, publication articles, and PMUT-based companies in the last five years. Looking at the growth trends, it is predicted that by 2025, there will be at least an addition of 30 new hotspots doing PMUTs with the total number reaching 100. As is visible from Figure 2, Asia will continue to be the lead continent doing PMUTs with at least 10 new institutes getting involved in PMUT research in the next 5 years from China. Articles will continue to increase monotonically with more than 300 research articles and more than 10 review articles published in the next 5 years. The increase in the number of new companies is unpredictable and complicated, but it is believed that there will be at least 3 more major investments worldwide in this direction. As all research does have growth, saturation, and decay phases, PMUTs do have their own and it is foreseen that research in this direction will hit saturation by the next 7 to 10 years, thereby following decay.

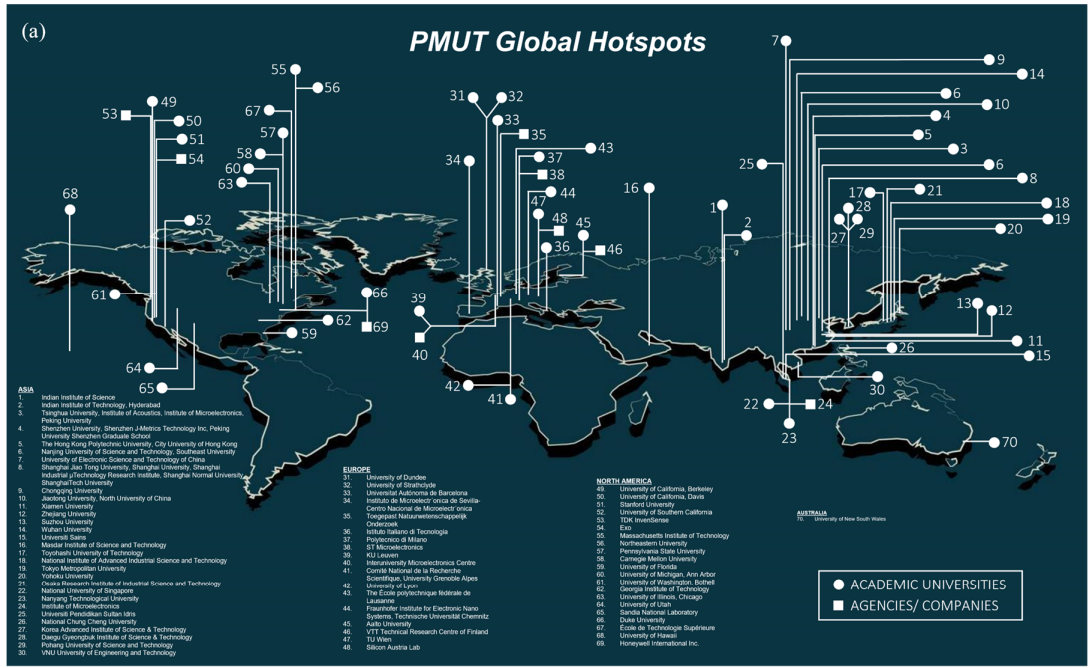
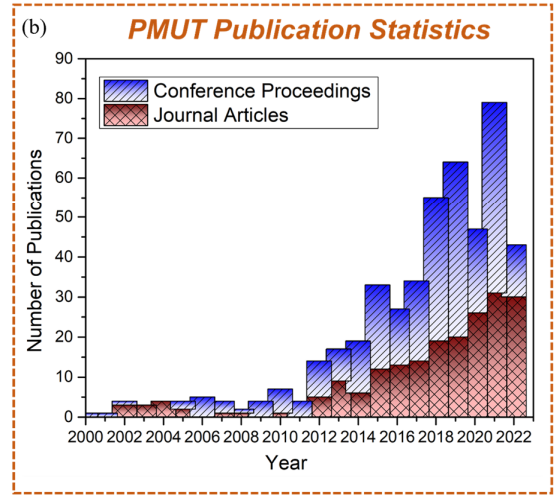
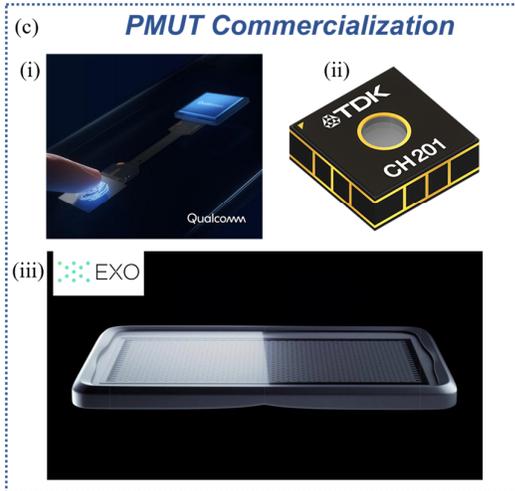

Figure 2. PMUT global development scenario (a) PMUT hotspots showing the active involvement of 4 continents – Asia, Europe, North America, and Australia (b) PMUT publication statistics from 2001 till 2022 (c) Major companies commercializing PMUTs (i) The 3D sonic sensor used as an in-display fingerprint sensor developed by Qualcomm Inc. (ii) The CH201, a 5 m PMUT rangefinder commercialized by TDK Inc. (iii) Cello, a handheld portable PMUT based imaging probe developed by Exo Inc.

## PMUT Design Basics
### PMUT Structural Classification
**Structural representation of PMUT**

PMUTs vibrate out-of-plane to transmit/receive sound. At their fundamental frequency of vibration, the vibrating shape is best captured by a modified parabolic shape (Figure 3(a)). Structurally, a PMUT is comprised of two important layers – the device layer and the piezoelectric-active layer (Figure 3(b)). The device layer contains the neutral plane with centroid of each layer at a distance of $z_i$ from the neutral plane. Each layer has thickness, radius, elasticity, density, and Poisson's ratio of $h_i$, $a$, $E_i$, $\sigma_i$, $\rho_i$ and $\nu_i$ respectively.

**Basis for structural classification: the 'kappa squared'**

Structurally, PMUTs can behave as a plate or membrane, or both. A plate's vibroacoustic response is determined by flexural rigidity (Figure 3(c)) whereas a membrane's is by pre-tension (Figure 3(d)). This behavior is determined by a non-dimensional ratio of the product of net structural pretension ($T_e$) and squared of $a$ to the equivalent flexural rigidity ($D_e$) of the PMUT [32].

$$\kappa^2 = \frac{T_e a^2}{D_e}$$

where,

$$T_e = \sum \sigma_i h_i$$

$$D_e = \sum \left(\frac{E_i}{1-\nu_i^2}\right)\left(\frac{h_i^3}{12} + z_i^2 h_i\right)$$

This expression indicates that a small, thick PMUT and a big, thin PMUT will demonstrate plate behavior and membrane behavior respectively at constant $T_e$ [32]. Figure 3(e) presents the dependence of the nondimensional frequency parameter ($\alpha_{00}\beta_{00}$) for the first mode of vibration (obtained while solving PMUT's natural response) on $\kappa^2$, is characterized by three zones. In the first zone, $\alpha_{00}\beta_{00}$ is constant with $\kappa^2$ (0.1 to 1) demonstrating the plate behavior. Above 100, $\alpha_{00}\beta_{00}$ is linear with $\kappa^2$ characterizing the membrane behavior. From 1 to 100, mixed behavior is observed. Figure 3(f) depicts the dependence of maximum resonant deflection at the first mode of vibration ($w_0$) with $a$, with a 1.62 µm thickness and a constant $T_e$. Three zones are again observed with a plate, plate-membrane, and membrane divisions respectively.

## Working principle of the PMUTs: the plate and the membrane

The working principles of plate and membrane PMUTs are different. For a plate, applying DC voltage across the piezoelectric layer tends to strain it due to the $d_{31}$ piezoelectric effect. This strain is then restricted by the underneath device layer leading to in-plane normal stress resultants '$R$' working from the piezoelectric layer's centroid. Since plate PMUTs are thick, the neutral axis rests in the device layer (Figure 3(c)). The difference (lever arm) between the piezoelectric layer's centroid and the neutral axis, along with $R$ results in a bending moment $M$ bending the structure out-of-plane. Applying an AC voltage thus makes the structure vibrate. Alternatively, a membrane is already stressed due to the presence of a net pre-tension in the structure which coupled with structural in-homogeneity causes the PMUT to bend even without an electric field (Figure 3(d)). A DC voltage induces a certain level of strain ($\varepsilon$) in the structure, which may be due to the comparable thickness of the device and piezoelectric layer or due to a bigger size-to-stack thickness ratio leading to a change in the level of tension ($\tau$) in the structure, thereby changing the amount of bending. Applying an AC voltage makes the structure vibrate.

## PMUT Functional Classification
### Lumped Model of a PMUT

Functionally, a PMUT can be classified into three groups: transmitter, receiver, and transceiver, and a system-level model explains them best. A lumped parameter model has been developed by Dangi et al. [32] coupling the electrical domain with the mechano-acoustic domain with an ideal transformer which represents the piezoelectric electromechanical coupling (Figure 3(g)). The flow variable in the mechano-acoustic domain is the flow rate of the fluid volume displaced, represented as $q$. PMUT is a capacitor having capacitance $C_s$. $V_{in}$ and $P_{in}$ is the input driving voltage and pressure respectively and $P_{out}$ is the output pressure. $C_m$, $R_m$ and $L_m$ are the structural compliance, damping, and equivalent mass of the PMUT respectively. $\phi$ is the turns ratio, representing the coupling between the electrical and the mechano-acoustic domains.

### PMUT as a transmitter

Design maps are available [32] in size-tension subspace and along thickness. In size-tension subspace, it is observed that a PMUT's central deflection ($w_0$) is proportional to the square of $a$ in the plate regime and is almost independent of $a$ in the membrane regime (Figure 3(h)). Also, the iso-frequency lines (Figure 3(h)) (black-dashed curves) suggest that low residual tension devices have higher deflection sensitivity even if they have to be made smaller to maintain the same operating frequency. Along thickness, the PMUT is idealized of being composed of the passive ($h_{passive}$) and device layers to add up to $h_{sum}$. The maximum transmitted pressure ($P_{TX}$) is plotted with respect to the thickness ratio $h_{passive}/h_{sum}$ (at $a$ = 250 µm; $T_e$ = 0 N/m). It has been reported that the optimal $h_{passive}/h_{sum}$ is 0.6 with pressure increasing with the increase in overall PMUT's thickness (Figure 3(i)).

### PMUT as a receiver

PMUT as receiver generates charge ($Q_{rx}$) which is directly proportional to $w_0$ [32]. In size-tension subspace, both $w_0$ and $Q_{rx}$ increases as $a^4$ (if $T_e$ = 0) and as $a^2$ if the PMUT is tension dominated (Figure 3(j)). Additionally, it depicts that the receiver at a lower operational frequency will have higher deflection and charge output for a given layer configuration. In the thickness subspace, $Q_{rx}$ is plotted with respect to $h_{passive}/h_{sum}$ (Figure 3(k)). It is observed that $Q_{rx}$ is directly proportional to $h_{passive}/h_{sum}$ and is inversely proportional to the thickness of the piezoelectric layer for a given $h_{sum}$. The plot also depicts that $Q_{rx}$ decreases with $h_{sum}$, which might be due to enhancement in the stiffness of the overall structure thereby leading to smaller deflections and smaller piezoelectric strain.

### PMUT as a transceiver

The design map for a transceiver PMUT can be obtained by combining the effects of transmitter and receiver [32]. Figure 3(l) represents the dependence of $Q_{rx}$ on $h_{passive}/h_{sum}$ when the receiver is kept 1 m away from the transmitter. The optimal thickness ratio to achieve the maximum $Q_{rx}$ is reported to be 0.7.

## PMUT Fabrication
### Thin-film Piezoelectric Materials

PMUTs have been fabricated with several piezoelectric thin-films, with a popular contribution from Muralt et al. [33] where transducers' figure of merit from thin-film materials perspective is mentioned. Some important materials are discussed as follows. with some important material properties tabulated in Table I. The first thin-film piezoelectric material which was developed for PMUTs was the planar magnetron sputtered ZnO [34], [35] (Figure 4(a(i))) which are reported to have excellent quality in terms of c-axis orientation with an X-ray rocking curve FWHM lesser than 1°. The next popular ferroelectric material developed for PMUTs is the PZT (Figure 4(a(ii))). It is observed that PZT has the highest piezoelectric constant ($|e_{31,f}| \sim 12$ C/m$^2$) [4], thereby working best as transmitters. However, simultaneously, PZT suffers from a high dielectric constant ($\varepsilon_{33,f} \sim 1200$) making it unsuitable for making receivers. Next, P(VDF-TrFE) [36] was developed (Figure 4(a(iii))) having the advantages of flexibility/stretchability and low-temperature process compatibility. The general

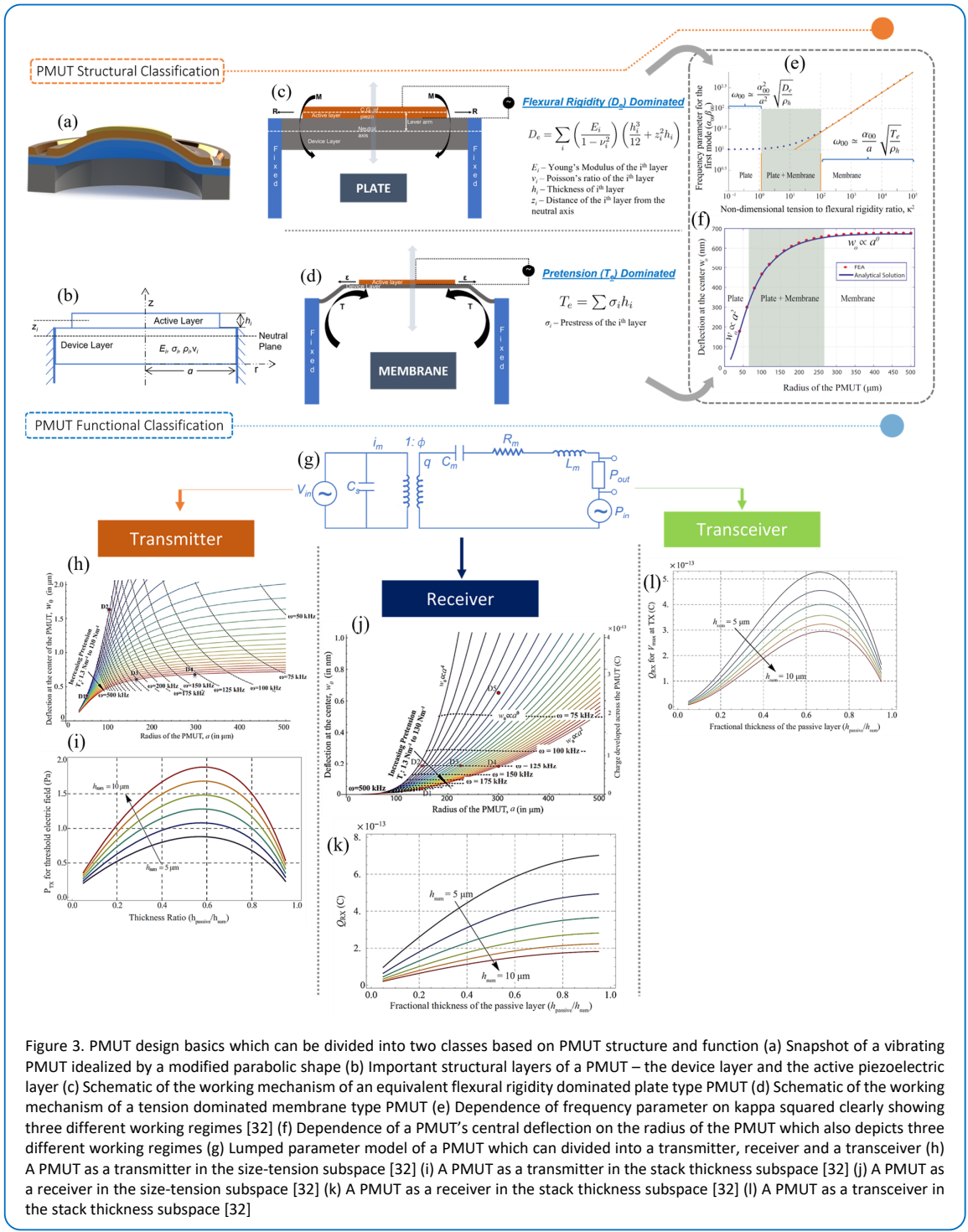

Figure 3. PMUT design basics which can be divided into two classes based on PMUT structure and function (a) Snapshot of a vibrating PMUT idealized by a modified parabolic shape (b) Important structural layers of a PMUT – the device layer and the active piezoelectric layer (c) Schematic of the working mechanism of an equivalent flexural rigidity dominated plate type PMUT (d) Schematic of the working mechanism of a tension dominated membrane type PMUT (e) Dependence of frequency parameter on kappa squared clearly showing three different working regimes [32] (f) Dependence of a PMUT's central deflection on the radius of the PMUT which also depicts three different working regimes (g) Lumped parameter model of a PMUT which can divided into a transmitter, receiver and a transceiver (h) A PMUT as a transmitter in the size-tension subspace [32] (i) A PMUT as a transmitter in the stack thickness subspace [32] (j) A PMUT as a receiver in the size-tension subspace [32] (k) A PMUT as a receiver in the stack thickness subspace [32] (l) A PMUT as a transceiver in the stack thickness subspace [32]

deposition technique involves CSD of 10-15 wt% of copolymer pellets dissolved in methyl-ethyl-ketone, followed by low-temperature annealing. The fourth material is AlN (Figure 4(a(iv))), which enabled PMUTs to be CMOS compatible, allowing monolithic integration of PMUTs with ASICs. Low-temperature sputtered AlN has a rocking curve having FWHM of 3°, $|e_{31,f}|$ of 1.05 C/m$^2$ and $\varepsilon_{33,f}$ of 10.5 respectively [18], [37], [38]. AlN thin-films are most suitable for PMUT receivers for their small $\varepsilon_{33,f}$. Next, Sc-AlN is developed with enhanced $|e_{31,f}|$ thereby improving a PMUT's overall performance. A 1 um Sc$_{0.15}$AlN film is developed by Wang et al. [26] (Figure 4(a(v))), having the rocking curve FWHM of 1.9° for a scandium doping of 15%. The $|e_{31,f}|$ has been deduced indirectly from the frequency response of a Sc-AlN PMUT and is found to be 1.6 C/m$^2$. Next, a PZT Fibered-Epitaxial thin-film was developed by Thao et al. [27] on oxide buffered layers by using magnetron sputtering followed by fast cooling (Figure 4(a(vi))). The thin-

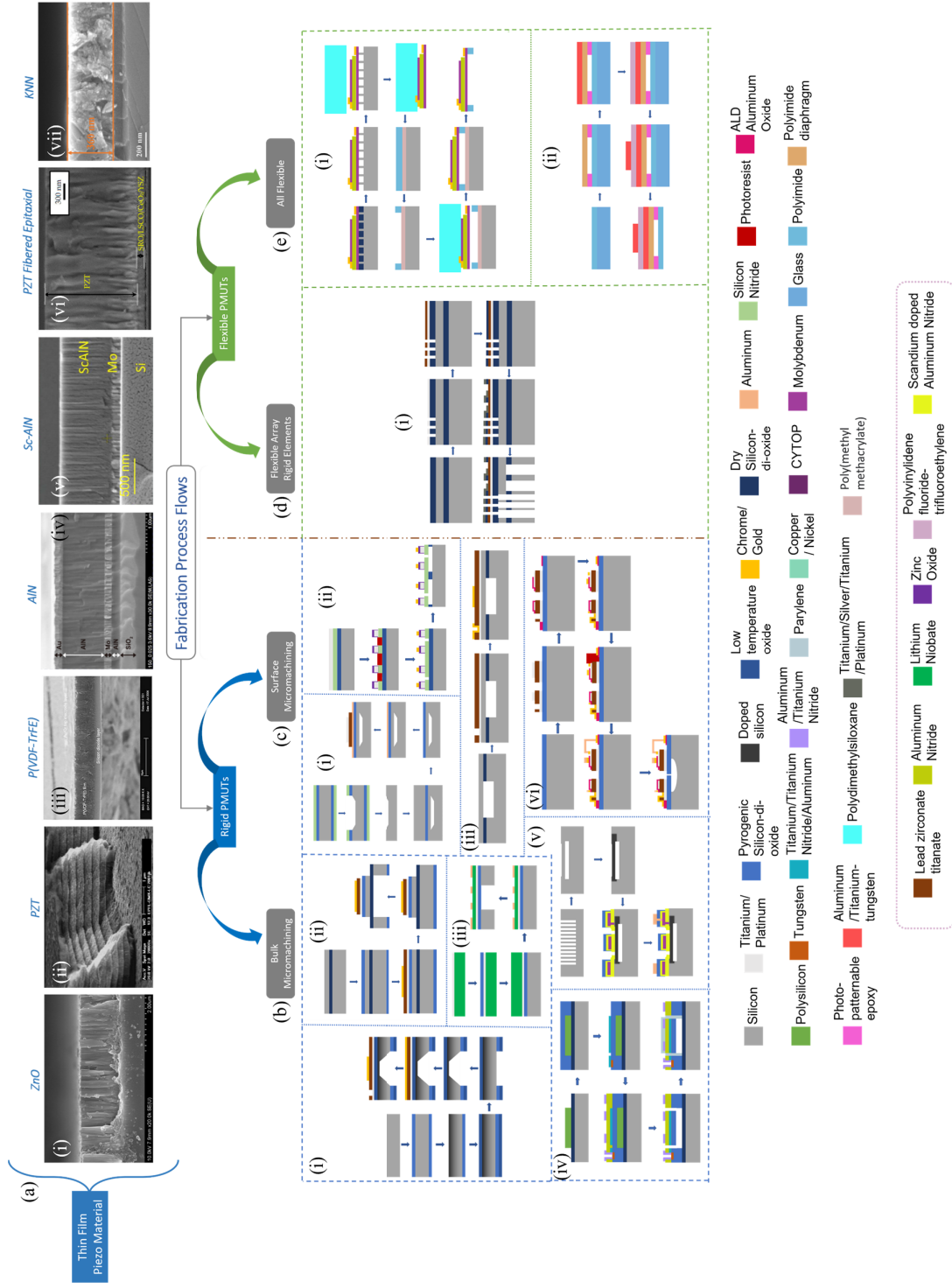

Figure 4. PMUT fabrication is divided into two parts – making of the piezoelectric thin-film material and fabricating the PMUT structure having the piezoelectric thin-film inside by nanofabrication tools (a) Various thin film piezoelectric material used till date (i) Zinc Oxide (ZnO) [34] (ii) Lead Zirconate Titanate (PZT) [4] (iii) Polyvinylidene fluoride-trifluoro ethylene P(VDF)-TrFE [36] (iv) Aluminum Nitride (AlN) [38] (v) 15 % Scandium doped AlN (Sc0.15AlN) [26] (vi) PZT fibered epitaxial film [27] (vii) Sodium Potassium Niobate (KNN) [41]. Fabrication methods divided into (b) PMUTs fabricated by bulk micromachining techniques (i) wet etching of bulk silicon (ii) dry etching of bulk silicon with a thinned-down bulk ceramic (c) PMUTs fabricated by surface micromachining techniques (i) prefabricated wet-etch holes followed by wafer bonding (ii) surface release by wet etch (iii) cavity SOI and surface machining (iv) surface release by polysilicon wet etch (v) surface micromachining by silicon migration (vi) surface release by wet etch (d) flexible PMUT array with rigid elements (i) fabrication of PMUT islands with connected springs (e) All flexible PMUTs (i) flexible PMUTs made by PDMS stamp (ii) flexible PMUT with flexible piezoelectric material on flexible substrate

Table I. Figures of merit for various thin film piezoelectric material used in PMUT

| Figures of Merit | Notation | PZT [4] | PZT Fibered Epitaxial [27] | Single Crystal PZT [39] | Epitaxial PMnN-PZT [40] | AlN [18],[37],[38] | $Sc_{0.15}AlN$ [26] | ZnO [34],[35] | KNN [41] |
|---|---|---|---|---|---|---|---|---|---|
| Deflection force, piezoelectric charge in deformed PMUT | $\|e_{31,f}\|$ (C/m$^2$) | 8 - 12 | 14 | 24 | 14 | 1.05 | 1.6 | 1 | 8.5 - 14.4 |
| Dielectric constant | $\varepsilon_{33,f}$ | 300 - 1300 | 200 - 300 | 308 | 250 | 10.5 | 12 | 10.9 | 445 |
| Piezoelectric charge in deformed PMUT | $\|e_{31,f}\|/\varepsilon_0\varepsilon_{33}$* (GV/m) | 0.7 - 1.8 | 5.3 - 7.9 | 8.8 | 6.3 | 11.3 | 15 | 10.3 | 2.2 - 3.6 |
| Coupling coefficient for flexural wave | $e_{31,f}^2/\varepsilon_0\varepsilon_{33}$* (GPa) | 5.6 - 54.2 | 110 - 74 | 211 | 88 | 11.9 | 24 | 10.3 | 18.3 - 53 |
| Dielectric loss angle | tan $\delta$ (@ 1 to 10 kHz, 10$^5$ V/m) | 0.01 -0.03 | - | - | - | 0.003 | - | 0.01 to 0.1 | - |
| Signal-to-noise ratio | $\|e_{31,f}\|/(\varepsilon_0\varepsilon_{33}$tan $\delta)^{0.5}$ (Pa$^{0.5}$)* | 4 - 8 | - | - | - | 20 | - | 3 to 10 | - |

*Relative permittivity of free space ($\varepsilon_0$): 8.85 pF/m film exhibited a $|e_{31,f}|$ of 10 to 14 C/m$^2$ and an $\varepsilon_{33,f}$ of 200 - 300 respectively. Next, single crystal thin-film PZT [39] and epitaxial thin-film PMnN-PZT [40] have been developed with a $|e_{31,f}|$ of 24 C/m$^2$ and 14 C/m$^2$ respectively. Recently, a sodium potassium niobate (KNN) (Figure 4(a(vii))) has been developed for making PMUTs [41]. KNN is lead-free and thus non-toxic and has better compatibility with CMOS. The layer was deposited following the repetitive CSD followed by sessions of pyrolysis and annealing treatments to achieve an overall film thickness of 360 nm. The KNN thin-film demonstrated a $|e_{31,f}|$ of 8.5 to 14.4 C/m$^2$ with a $\varepsilon_{33,f}$ of 445.

## Fabrication Process Flows

Fabrication-wise, a PMUT array can be broadly classified into rigid and flexible arrays. Rigid arrays have zero conformability and target applications that don't demand surface attachability whereas flexible arrays can be partial to fully conformable.

### Rigid PMUT Arrays
#### Bulk Micromachining

In this, the bulk substrate is etched by deep etching methods to release the PMUT's diaphragm. The first work [42] in this direction starts with a silicon wafer with pyrogenic oxide coated on the bottom surface which is followed by boron diffusion, deposition of low-temperature oxide, backside oxide etches, and deep silicon wet etching by EDP to release the diaphragm. The top surface was coated with metalized sol-gel PZT sandwich, which was then wet etched to establish contacts (Figure 4(b(i))). The second work (Figure 4(b(ii))) starts with a metalized and surface oxidized silicon-on-insulator (SOI) substrate which is followed by top metal patterning, PZT/bottom metal, oxide, and device silicon etching to define the top structure, followed by a deep reactive ion etching (DRIE) of the bulk silicon to release the membrane [43]. The third work [31] starts with a metalized bulk lithium niobate (LiNbO$_3$) crystal which was then wet oxidized. The assembly was then flipped and bonded to the surface of silicon and polished down to the thickness of thin-film. The stack was deposited with metal on top which was then patterned. A DRIE was performed from the back to release the membrane (Figure 4(b(iii))).

#### Surface Micromachining

In this, the diaphragm is released by micromachining from the surface without etching the bulk silicon. The first work [44] starts with a wet-oxidized silicon wafer that is coated with silicon nitride. A shallow trench is then defined in silicon by reactive ion etching (RIE). Another silicon wafer with an oxide coat facing downwards is then bonded to the previously machined wafer. The top silicon is then thinned to a couple of microns, followed by the PZT metal sandwich deposition, followed by the top metal and PZT etch (Figure 4(c(i))). The second work (Figure 4(c(ii))) starts with a silicon wafer that is oxidized (low-temperature oxide), coated with nitride, and metalized. ZnO is then sputtered and etched along with the bottom metal and nitride to the desired pattern. Photoresist (PR) is pattern deposited in the trenches and the resultant stack's top surface is metalized. The stack is next patterned to create etch holes in the nitride layer. The oxide is then wet-etched to release the membrane [45]. The third work begins with a cavity SOI wafer (Figure 4(c(iii))) which is metalized and coated with PZT followed by its etch to take the ground connection, followed by the deposition and patterning of the top metal for live connection respectively [46]. The fourth work (Figure 4(c(iv))) starts with a silicon wafer deposited with a layer of patterned polysilicon. It is then covered with oxide which is next coated with Titanium/Titanium-Nitride/Aluminum after partial etching and filling with tungsten. AlN is next sputtered and then coated with a pattern of Aluminum/Titanium-Nitride, followed by an etch and sputter step to establish ground contact. The stack is patterned, etched, and the polysilicon sacrificial etched to release the membrane. The assembly is then coated with Parylene-C to seal off the cavity [47]. The fifth work starts with [48] a silicon wafer with embedded shallow holes. The wafer is then annealed to activate silicon migration, which then forms a thin membrane over a cavity, forming a membrane. The membrane is next doped and deposited with patterned Sc-AlN and metal, followed by the pattern

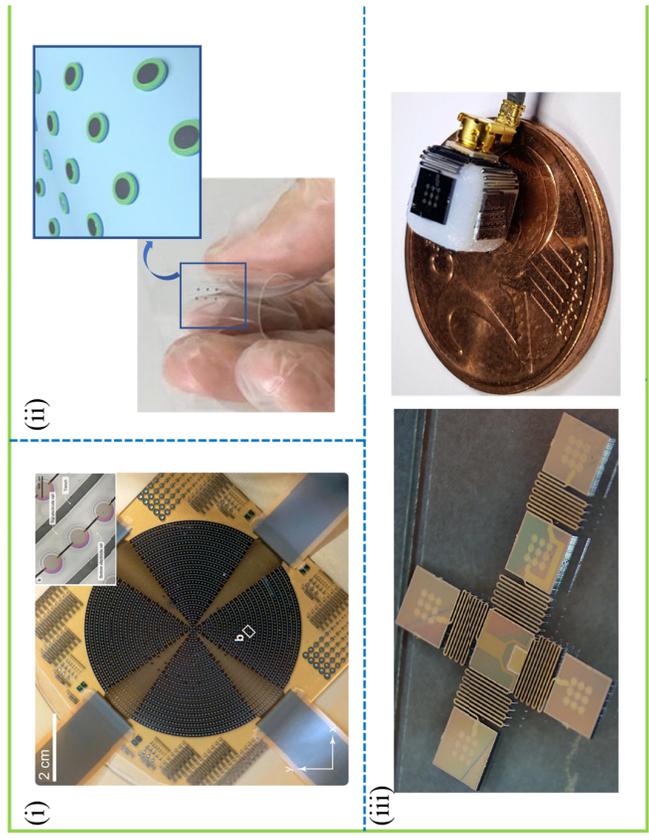
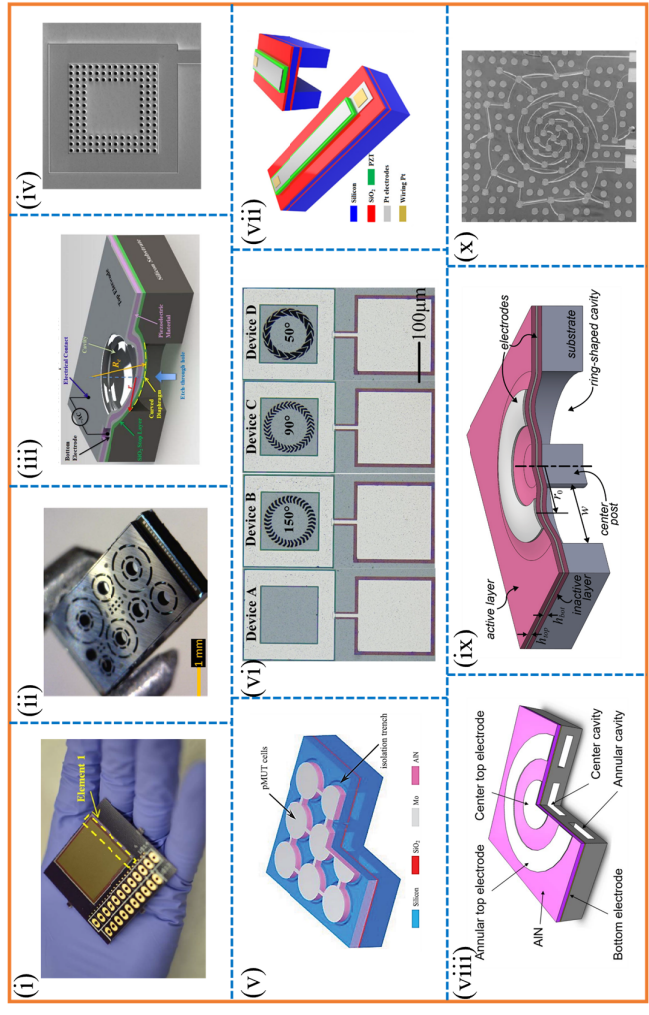

Figure 5. Special PMUTs which can be classified as (a) structurally modified PMUTs (i) bimorph PMUT with increased pressure output [54] (ii) PMUT with relaxed boundary condition [55] (iii) PMUT with isolation trench and higher sensitivity [58] (vi) PMUTs with v-shaped rings [59] (vii) Broadband PMUTs [60] (viii) central and annular PMUT on a single chip [63] (ix) ring-shaped PMUT [64] (x) Archimedean PMUT array [65] (b) flexible PMUTs (i) polymer PMUTs [70] (ii) fully flex PMUT [51] (iii) flexible PMUT die with rigid elements [50]

coat with oxide and metal. Finally, the released silicon membrane is pattern etched to define membranes of desired shape and dimension (Figure 4(c(v))). The sixth work [49] starts with an oxidized silicon wafer which is then coated with patterned metal PZT sandwich. Aluminum oxide is then pattern deposited as shown in Figure 4(c(vi)) which is followed by further metallization to define the in-device traces. PR is pattern coated between the top metal and the bond-pad which is then pattern sputtered with metal in order to establish the connection. The oxide attached to the silicon substrate is pattern etched followed by a surface isotropic etch of the silicon to release the diaphragm.

### Flexible PMUT Arrays
#### Flexible Array with Rigid Elements
In this category, the PMUT array is flexible while each PMUT element is rigid. Sadeghpour et al. [50] start with an oxidized SOI substrate which is metalized, and pattern etched in a desired fashion. PZT is next pattern deposited followed by further patterned oxidization and metallization to establish top contacts. A subsequent DRIE is followed to etch the bulk silicon to release the membrane followed by a patterned RIE to desire etch silicon from the top to create free serpentines (Figure 4(d(ii)).

#### All Flexible PMUTs
In this category, both PMUT elements and the array are flexible. The first work starts with a layered stack as shown in (Figure 4(e(i))). A PDMS stamp is then used to pull off the functional layers and transfer it to another stack having silicon, poly(methylmethacrylate), and patterned polyimide. Poly(methylmethacrylate) is then stripped off to detach and create the flexible PMUT [51]. The second work begins with a polyimide-on-glass substrate, which is then patterned with a photo-patternable epoxy (Figure 4(e(ii))). A polyimide diaphragm is suspended on top of it which is followed by the deposition of the metal PVDF sandwich. The stack is then detached from the glass substrate to create flexibility [29].

### Author's Sectional Forecast
Material-wise, there has been a huge development in finding new materials since the last decade with at least 5 new materials developed. Out of these, single crystal PZT will be most effectively used in transmitters having the highest piezoelectric coefficient, and Sc-AlN most effectively used in receivers having the highest material charge sensitivity. From the past trends, it is predicted that there will be at least 5 new materials developed in the next 5 years with better transmit-receive efficiencies, with a major focus on developing transparent and flexible thin films. Fabrication-wise, there have already been several developments in the past decade and not many newer contributions are envisioned at least toward making rigid PMUTs. However, as evident from Figure 4(e), there is a huge room for developing flexible/stretchable PMUT arrays and it is predicted that there will be more than 5 new flows developed in this direction in less than 2 years.

## Special PMUTs
This section describes the structurally and functionally nonconventional PMUTs and is broadly classified into – structurally modified PMUTs and Flexible PMUTs.

### Structurally Modified PMUTs
PMUTs have been structurally modified to enhance their performance in terms of deflection/transmit sensitivity, directivity, bandwidth, etc. Akhbari et al. fabricated bimorph AlN PMUTs having two piezoelectric layers in an array as shown in Figure 5(a(i)). They claim that their PMUTs have four times the electromechanical coupling coefficient as compared to unimorph AlN PMUTs of similar geometry and frequency [52]. Rozen et al. have fabricated PMUTs with venting rings (Figure 5(a(ii))) to amplify the far-field sound pressure level (SPL). An increase of 4.5 dB over a control device has been claimed [53]. Akhbari et al. have developed curved PMUT (Figure 5(a(iii))) with a radius of curvature 400-2000 µm having 50 times the deflection sensitivity than a comparable flat device [54]. Wang et al. etched holes along the periphery of PMUT (Figure 5(a(iv))) which increased the SPL by 5.3 dB as compared to a similar device [55]–[57]. Wang et al. have created isolation trenches along the PMUT periphery (Figure 5(a(v))) which has increased the output pressure by 76 % than a similar device. The authors have claimed that the trench reduces the deflection-induced tensile membrane stress, allowing more motion [58]. Chen et al. have reported on V-shaped surface springs (Figure 5(a(vi))) which increase the deflection sensitivity by 203% due to the release of residual stress owing to a flat membrane [59]. Wang et al. has fabricated a wide frequency band rectangular PMUT by using the technique of mode merging (Figure 5(a(vii))). The -6 dB bandwidth has been reported to be 95% in water, which is exceptionally high compared to the control devices [60]–[62]. Wang et al. have fabricated a combination of central and annular PMUT-on-a-single-chip (Figure 5(a(viii))) which has increased the transmit and receive sensitivities by 1.9 and 6.5 times a control PMUT. They claim the observation to be caused due to the coupling between the two separate membranes [63]. Eovino et al. have reported on a ring-shaped PMUT (Figure 5(a(ix))) with a central post and have observed the velocity bandwidth in fluid-coupled operations to reach 160% which is claimed to be more than 60% of bandwidth ever reported. They claimed that the ring geometry and acoustic-induced resonance were the cause of the broadband nature [64]. Wang et al. have reported on a spiral Archimedean PMUT array (Figure 5(a(x))) which is claimed to generate 18% higher sound pressure as compared to conventional phased PMUT arrays of similar dimensions [65].

### Flexible PMUTs
There has been a recent surge in wearable transducers' development [66]–[69]. PMUTs have also been make-wise modified to develop flexible arrays having the added advantage of wearability and can conformability to various surface topologies. Chare et al. fabricated large-area PMUTs based on flexible PVDF (Figure 5(b(i))) and have demonstrated their use in flat panel display compatibleness. Using the phased array, they have created an acoustic twin trap, claiming to use their array for in-air haptics [70]. Although the array was fabricated on a glass substrate, the same group has reported on fully flexible arrays as well [29]. Sun et al. have reported on a PMUT array sticker (Figure 5(b(ii))) by using a PDMS stamp-based transfer. The devices have a resonant frequency and displacement sensitivity of 2.58 MHz and 30 nm/V respectively [51]. Sadeghpour et al.

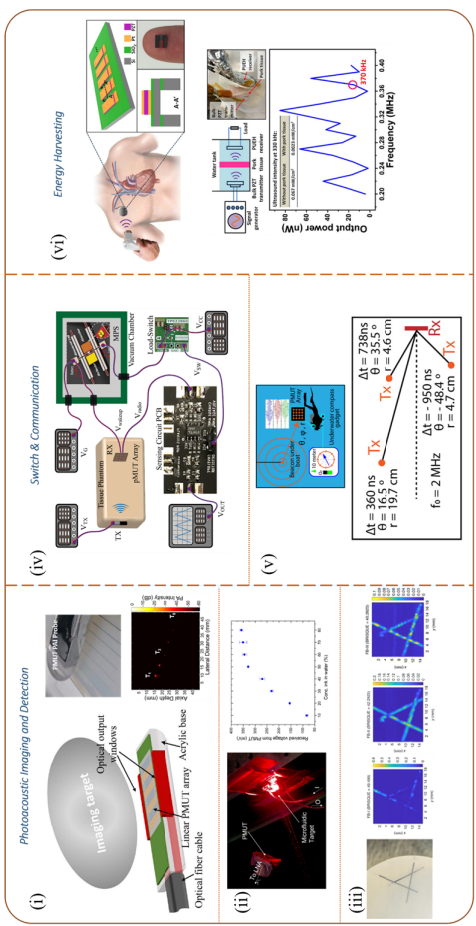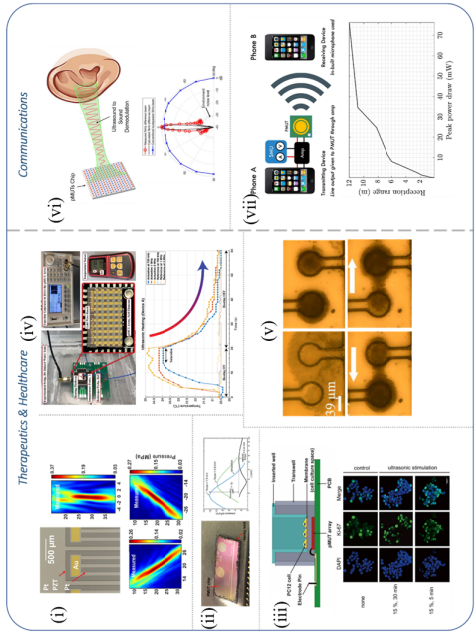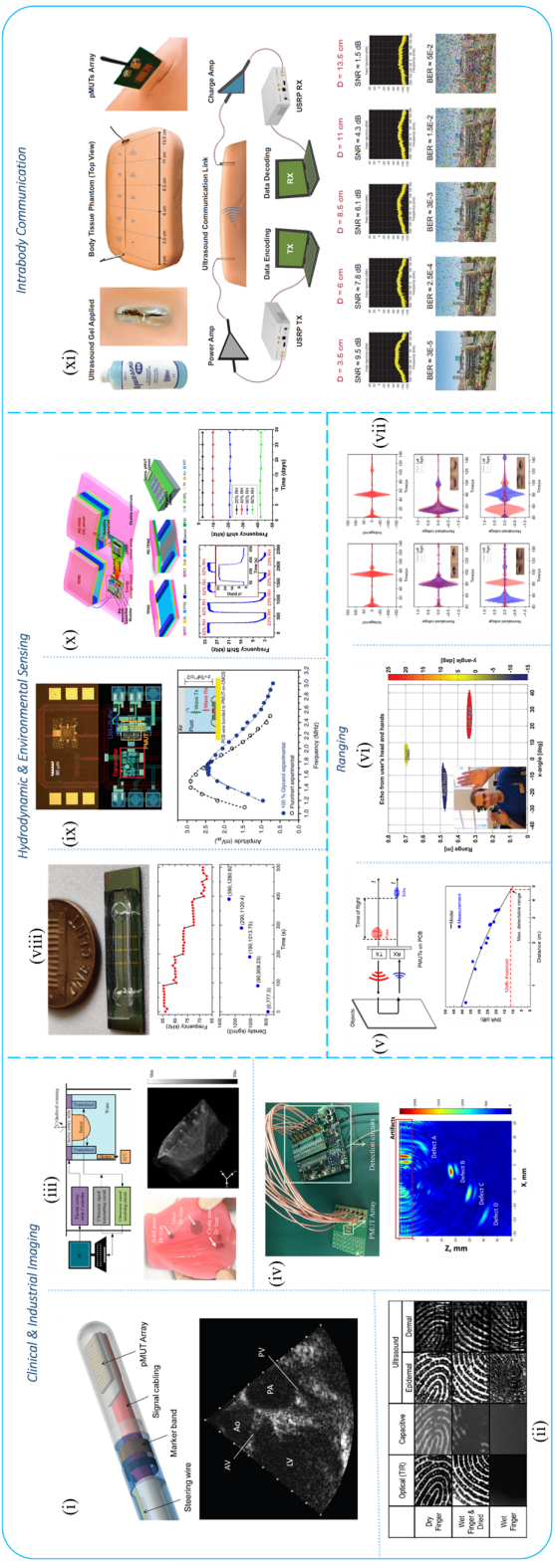

Figure 6. PMUT Applications as (a) transmitters (i) 32 element linear phased array for neuromodulation [71] (ii) annular phased array for neuromodulation application [74] (v) PMUTs applied to acoustofluidics for particle streaming [75] (vi) PMUTs for directional ultrasound [76] (vii) near ultrasound PMUT for in-air communication [77] (b) PMUTs as receivers (i) photoacoustic imaging system using PMUTs [79] (ii) photoacoustic spectroscopy system for concentration detection [81] (iii) PMUTs operating at higher modes for photoacoustic imaging application [82] (iv) PMUTs as zero powered wake-up switches for communication [83] (v) Source localization system using PMUTs [84] (vi) broadband energy harvester using PMUT [25] (c) PMUTs as transceivers (i) IVUS intracardiac imaging system using high density PMUT arrays [88] (ii) 3D ultrasound on chip by using PMUT with ASICs [89] (iii) rotational tomography system by using PMUT [90] (iv) PMUTs for imaging defects in solids [91] (v) long range finder using single crystal PZT based PMUTs [39] (vi) In air gesture recognition system by using AIN PMUTs [92] (vii) eye blinking system by using PMUTs [93] (viii) PMUT microfluidic integration for fluid density sensing [98] (ix) single cell PMUT on a chip for hydrodynamic property sensing [100] (x) TENG powered PMUTs for gas sensing applications [28] (xi) PMUTs for intrabody communication [31]

have created bendable PMUT arrays (Figure 5(b(iii))) connected by silicon springs and have demonstrated the array's capability by wrapping around a 3D cube having 5 mm dimension, with a 90° bendability [50].

## PMUT Applications

PMUT applications can be classified as transmitter, receiver, and transceiver as described below and some important comparisons in such terms have been tabulated in Table II.

## PMUTs as Transmitters

As transmitters, PMUTs generate sound and can be applied to – therapeutics and healthcare, and communications. In the first direction, Tipsawat et al. [71] developed a 32-element phased array PMUT with beam steering using Nb-doped PZT for neuromodulation (Figure 6(a(i))). An acoustic pressure of 0.44 MPa has been obtained at a focal distance of 20 mm with an average intensity of 1.29 W/cm$^2$. Eovino et al. have reported on concentric PMUT arrays [72] for focused ultrasound applications could focus on a spot of 1.9 mm for a pressure of 12 kPa/V (Figure 6(a(ii))). Next, Lee et al. has created a low-intensity pulsed ultrasound system [73] by using a PMUT linear array which generated a pressure of 0.15 MPa at 1 mm and increased cell proliferation rate in the range of 138 – 166% with respect to the control condition (Figure 6(a(iii))). Pop et al. have created a bio-heating platform [74] with a 5x10 PMUT phased array and demonstrated an increase in 4°C relative temperature in 10 s (Figure 6(a(iv))). In acoustofluidics, Cheng et al. [75] have developed confined PMUT arrays to trap and manipulate 4 µm silica beads using unipolar excitation (Figure 6(a(v))). In communications, Shao et al. [76] reported on a parametric air-coupled single-chip bimorph PMUT array to generate highly directional audible sound with half power beam width lesser than 5°. They demonstrated generating a 5 kHz audio by combining frequencies of 252 and 257 kHz respectively (Figure 6(a(vi))). Harshvardhan et al. [77],[78] have developed near ultrasound PMUTs for sending data over sound. It has been demonstrated that PMUTs can send data successfully for a range of 6 m, while consuming less than 10 mW electrical power Figure 5(a(vii)).

## PMUTs as Receivers

As receivers, PMUTs receive sound and can be applied to – photoacoustic imaging and detection, switch and communication, and energy harvesting. In photoacoustic imaging, Dangi et al. [79],[80] has combined PMUT array with a pulsed light source in a single device which could image a custom-made phantom with lead targets embedded as shown in Figure 6(b(i)). The PMUTs in the device had a center frequency of 6.75 MHz in water with a photoacoustic bandwidth of 89%. Next, Roy et al. [81] developed a PMUT-based pulsed photoacoustics microfluidic concentration detector capable of detecting ink-water concentrations (see Figure 6(b(ii))). Recently, Cai et al. [82] made a high-order multi-band AlN PMUT and used PMUT's higher modes to get better photoacoustic image from a custom-made phantom (Figure 6(b(iii))). They used BRISQUE as an image quantifier and observed the higher modes of PMUTs to give a better image as compared to the lower modes at similar depths. In switches and communication, Pop et al. [83] developed a zero-power PMUT-based ultrasonic wake-up receiver, a zero-power MEMS plasmonic switch, and a low leakage current CMOS load-switch as important parts of the system. A working demonstration of the wake-up behavior has been shown in a range of 5 cm (Figure 6(b(iv))). Herrera et al. [84] have made PMUT based underwater acoustic source localization system, based on the difference in time of arrival of ultrasound bursts from a source which made it possible to detect three different sources at different locations (Figure 6(b(v))). In the domain of energy harvesting [85]–[87], Shi et al. [25] has demonstrated making rectangular PMUT-based broadband energy harvester for self-powered IMDs (Figure 6(b(vi))). They demonstrated power transfer through a 6 mm pork tissue in water to receive a power of 2.3 µW/cm$^2$.

## PMUTs as Transceivers

As transceivers, PMUTs transmit and receive sound, can function with a differential actuation-sensing scheme as frequency-shift devices, or can function as a transmit-receive pair with device twins facing each other. In the first direction, Dausch et al. [88] fabricated two 256 and 512-element arrays through silicon vias to make an intracardiac catheter which captured a B-mode image in-vivo from a porcine model (Figure 6(c(i))). Next, Tang et al. [89] proposed a 3-D ultrasonic fingerprint sensor on-chip consisting of a 110x56 PMUT array bonded to a custom 180 nm readout ASIC and can image epidermal, subsurface, and dermal fingerprints (Figure 6(c(ii))). Next, Liu et al. [90] constructed a rotational tomography setup by using 4 (1x128) PMUT linear arrays which generated clear images of the phantom (Figure 6(c(iii))). Next, Ji et al. [91] devised a 3 MHz, 16-channel PMUT array-based ultrasonic imager which was imaged defects inside a solid from a depth of 3 cm (Figure 6(c(iv))). In ranging, Luo et al. [39] created an airborne 40 kHz PMUT for long-range detection of 4.8 m. The design used two different PZT PMUTs having a high piezoelectric coefficient ($e_{31,f}$ ~ 24 C/m$^2$), one as a transmitter, the other as a receiver (Figure 6(c(v))). Next, Przybyla et al. [92] made a 3D on-chip rangefinder to localize targets over a 90° field of view till a 1 m distance (Figure 6(c(vi))). Sun et al. [93] worked on a portable eye-blinking monitoring application by mounting a similar PMUT on a spectacle that could track eye blinks (Figure 6(c(vii))). Next, as frequency shift applications [94]–[97] Roy et al. devised a PMUT-microfluidic integrated device [98], [99] to sense fluid density and demonstrated real-time density monitoring with the device [30] (Figure 6(c(viii))). Ledesma et al. [100] have developed a single-cell PMUT on a 130 nm CMOS chip to monitor fluid properties such as density, viscosity, and sound compressibility (Figure 6(c(ix))). Next, Sun et al. [28] constructed TENG powered functionalized PMUT to demonstrate its working as a combined temperature and humidity sensor (Figure 6(c(x))). As transmit-receive pair, PMUTs have found application in intrabody communication, in which Pop et al. [31] created a special in-plane actuated PMUT using anisotropic LiNbO$_3$ thin-film for enhanced bandwidth. Image data was sent over a distance of 13.5 cm through body tissue phantom (Figure 6(c(xi))).

## Author's Sectional Forecast

As transmitters, there has been an increasing trend to develop focused ultrasound applications with PMUTs and it is believed that the world will witness at least 7-10 more such applications with different PMUT configurations happening in the next 5 years. As receivers, there has been a huge development in the domain of photoacoustics in the past 7 years, and it is felt that there is a huge scope for investment in this direction. It is expected to witness more than 15 different applications in less than 5 years in this direction. As transceivers, ultrasound ranging, and imaging will continue to dominate with more than 15 contributions in the next 5 years along with an increasing contribution in the domain of hydrodynamic and environmental sensing.

Table II. Application classified PMUT performance comparison

| Application category | Piezoelectric material | PMUT geometry | PMUT lateral dimension† | PMUT stack thickness (st + ac) (μm)* | Array configuration | Operating frequency | Deflection sensitivity | Transmit Sensitivity | Receive/ Sensing Sensitivity | BW# @ -6 dB | Applied for |
|---|---|---|---|---|---|---|---|---|---|---|---|
| Transmitter | PZT | Rectangular | (l x w): 8.3 mm x 680 μm | 5 + 1.5 | 32 channels, linear array | 1.4 MHz in water | - | 45.9 kPa/V, @ 20 mm (focused) | - | - | Phased array for neuromodulation [71] |
| | AlN | Annular, concentric | w: 60 μm | 5 + 0.83 | 5 channels | 6 MHz in mineral oil | - | 12 kPa/V @ 1.9 mm (focused) | - | - | Proposed phased array for catheter HIFU [72] |
| | PZT | Circular | a: 120 μm | 3.7 + 1 | 10 x 29 channels | 1.5 MHz in culture medium | - | 30 kPa/V @ 1 mm (unfocused) | - | - | LIPUS for cell stimulation [73] |
| | PZT | Circular | a: 60 μm | 4 + 1.9 | - | 8 MHz in water | 40 nm/V @ 6.8 MHz | 1.9 kPa/V @ 7.5 mm (unfocused) | - | 62.5% | Particle manipulation [75] |
| Receiver | PZT | Circular | a: 60 μm | 4 + 1.9 | 65 channels, Linear array | 6.75 MHz in water | - | 3.24 kPa/V @ 7.5 mm (unfocused) | 0.48 mV/kPa @ 0dB gain | 89% | Photoacoustic imaging [79] |
| | AlN | Circular | a: 92 μm | 1.45 + 1 | 10 x 10 channels | 700 kHz in air | - | - | - | - | Zero power ultrasound receiver [83] |
| | PZT | Rectangular | (l x w): 500 μm x 250 μm | 11 + 2 | Single channel | 285 kHz in water | - | - | 2 mV/mW/cm$^2$ | 74% | Energy harvesting [25] |
| | PZT | Rectangular | (l x w): 110 μm x 80 μm | 6 + 1 | 512 channels, 32 x 16 | 5 MHz in tissue | - | - | - | 30% | Intravascular ultrasound [88] |
| Transceiver | AlN | Rectangular | (l x w): 58 μm x 43 μm | 1.2 + 1 | 6160 channels, 110 x 56 | 14 MHz in air | - | 625 Pa/V @ 250 μm (focused) | 344 μV/kPa | 33% in PDMS | 3D fingerprint sensor [89] |
| | PZT | Circular | a: 750 μm | 10 + 0.5 | 1 channel | 235 kHz in air | 1.5 μm/V | - | 0.56 mV/μm | - | Fluid density sensor [30] |
| | Single-crystal PZT | Circular | a: 1250 μm | 6.25 + 2 | 4 channel | 40 – 50 kHz in air | 10.75 μm/V | 0.4 Pa/V @ 33 cm (unfocused) | 2.8 mV/Pa | 11% | Long rangefinder [39] |
| | LiNbO$_3$ | Rectangular | (l x w): 170 μm x 130 μm | 1.35 + 1 | 1 channel, 15 x 15 | 630 kHz in water | 80 nm/V in air | - | - | 64% | Intrabody communication [31] |

†l: length, w: width, a: radius; *st: structural layer thickness, ac: active layer thickness; #BW: Bandwidth

## Conclusion

This concise review thus forms an all-in-one reference for the important works done since the dawn of PMUT making and serves to provide the readers with an overall awareness of PMUTs, their past and present, their design, and their potential for various applications. Although PMUTs bring the new ultrasound revolution, like any other technologies they have their own pros and cons. The pros have been discussed earlier. Some of the cons over traditional piezoelectric transducers are – the limitation to creating high-intensity ultrasound, relatively complicated backend, and bulky investment in terms of capital and time for new ventures. Despite these limitations, it is believed that like all other technologies, PMUTs will continue to survive and prosper in years to come and will be applied by fellow researchers and entrepreneurs to develop meaningful applications suitable to drive the MEMS ultrasound revolution.

## Acknowledgement

This research is supported by A*STAR under the "Nanosystems at the Edge" programme (Grant No. A18A4b0055).